\documentclass[12pt]{article}
\usepackage{graphicx}

\begin{document}

\centerline {\large \bf $\overline{p}p$ and $pp$ Elastic
Scattering in a Multipole-Pomeron Model}

\vskip .6cm

\centerline{K. Kontros$^{\ast}$, A. Lengyel$^{\dagger}$ and Z.
Tarics$^{\ddagger}$}

\vskip .3cm

\centerline{\sl Institute of Electron Physics, Universitetska 21,}
\centerline{\sl 88000 Uzhgorod, Ukraine}

\vskip .6cm

\begin{abstract}
We assume that the Pomeron is a sum of Regge multipoles, each
corresponding to a finite gluon ladder. From a fit to the
diffraction cone data of $pp-$ and $\overline{p}p-$ scattering we
found that the triple pole is significant for the rise of the
ratio $\sigma_{el}$/$\sigma_{tot}$ at high energies.
\end{abstract}

\section{Introduction}

It has been conjectured recently \cite{FJKLPP} that the Pomeron
instead of being an infinite gluon ladder \cite{BFKL} may appear
as a finite sum of gluon ladders corresponding to a finite sum of
Regge multipoles with increasing multiplicities. The first term in
the $\ln s$ series contributes to the total cross-section with a
constant term and can be associated with a simple pole, the second
one (double pole) goes as $\ln s$, the third one (triple pole) as
$\ln^2s$, etc. All Pomeron poles have unit intercepts. A new prong
opens each time the available energy exceeds the threshold value.
The values of threshold parameter were found \cite{FJKLPP} from a
fit to the experimental data. Another important set of parameters
is related to the coupling of the gluons to hadrons, giving the
relative weight of various diagrams. In ref. \cite{FJKLPP} they
were fitted to the forward scattering data and subsequently they
were also calculated in QCD. Below we study this problem both in
the forward and non-forward directions. Because of its complexity,
we do not consider here the effect of the progressively opening
channels as it was done in ref. \cite{FJKLPP}. Instead we first
consider a model with a simple and dipole Pomeron [DP]
contribution used earlier to fit elastic hadron scattering
\cite{VJS}, low-$x$ structure functions \cite{JPP} and
photoproduction of vector mesons at HERA \cite{FJP}.

The DP ansatz reads \cite{VJS}
\begin{equation}
\label{eq1}P(s,t)=isg_0\sum_{i=1}^2c_iR_i^2(s)e^{R_i^2(s)t},
\end{equation}
where $c_1=1$, $c_2=\lambda {b}-1=-\varepsilon $;
$R_1^2$$(s)=\alpha ^{\prime }(b+\ln (-i\frac s{s_0}))$;
$R_2^2$$(s)=\alpha ^{\prime }\ln (-i\frac s{s_0})$ . Apart from
the normalization factor $g_0$, this model contains 4 adjustable
parameters: $\lambda$, $b$, $s_0$ and $\alpha ^{\prime}$, moreover
their number can be still reduced within the cone region (i.e.
$|t|\leq 1$ $GeV^2$ by setting $\lambda =\frac 1b$). Here a linear
Pomeron trajectory, $\alpha \left(t\right) =1+\alpha
^{^{\prime}}t$ is implied. Generalization to include nonlinear
trajectory is straightforward. The model is strongly constrained
by an integral relation between the residue at the simple and
double poles \cite{VJS}, that produces the observed dip-bump
structure in the differential cross-section in accordance with the
experimental data.

\vskip .5cm

\hrule

\vskip .5cm

\noindent \vfill $ \begin{array}{ll} ^{\ast}\mbox{{\it e-mail
address:}} &
 \mbox{jeno@kontr.uzhgorod.ua}
\end{array}
$

$ \begin{array}{ll} ^{\dagger}\mbox{{\it e-mail address:}} &
 \mbox{sasha@len.uzhgorod.ua}
\end{array}
$

$ \begin{array}{ll} ^{\ddagger}\mbox{{\it e-mail address:}} &
   \mbox{iep@iep.uzhgorod.ua}
\end{array}
$

The Dipole Pomeron model with a unit Pomeron intercept \cite{VJ}
was used to describe successfully $pp-$ and $\overline{p}p-$
elastic scattering at the ISR and the Collider energies, however
the ratio $\sigma _{el}/\sigma _{tot}$ was found \cite{VJS} to
decrease asymptotically for all physical values of the parameters.
To obtain an increasing function for this ratio the authors of
\cite{VJS} introduced a factor corresponding to the supercritical
Pomeron behavior \cite{DL}. However in this case the unitarity is
broken. It was shown \cite{CMG} that the unitarity violation
occurs at energies only slightly above the Tevatron energy of 1.8
TeV, and therefore it is a problem of the present and not of the
future. To avoid this problem, we consider a model containing a
finite series of Pomeron terms up to $\ln^2s$, in accordance with
the unitarity constrains.

\section{The model}

Our ansatz for the scattering amplitude is:
\begin{equation}
\label{eq2}A_{pp}^{\overline{p}p}=P+R_f\pm R_\omega,
\end{equation}
where we introduce a tripole contribution to the DP in the
simplest way (see also \cite{DGJ}):

\begin{equation}
\label{eq3}P\left(s,t\right)=isg_0\left[b+\ln \left(-i\frac
s{s_0}\right)+c\ln^2\left(-i\frac s{s_0}\right)\right]
e^{\alpha'bt} e^{ \left(\alpha_{P}(t)-1\right)
\ln(-\frac{is}{s_0})}.
\end{equation}

The Pomeron trajectory:
\begin{equation}
\label{eq4}
\alpha_{P}(t)=\alpha_{P}(0)+\alpha_{P}'t+\gamma_{P}\left(\sqrt{t_{P}}-\sqrt{{t_{P}}-{t}}\right),
\end{equation}
where the two-pion threshold $t_{P}=4m_{\pi}^{2}$.

In (2) the $R_f,R_\omega $ contains the subleading Reggeon
contributions ($f$ and $\omega $) to the scattering amplitude:

\begin{equation}
\label{eq5} Rj\left( s,t\right) =g_j\left( -i\frac s{s_0}\right)
^{\alpha _j\left( t\right) }e^{b_jt},\alpha _j\left( t\right)
=1+\alpha _j^{^{\prime }}t,j=f,\omega; s_{0}=1 GeV^{2}.
\end{equation}

Recently in paper \cite{gay} the contribution of truncated BFKL
Pomeron series to the $\sigma _{tot}$ of $pp-$ and
$\overline{p}p-$ scattering was studied and it was shown that a
reliable description can be obtained by using two orders in this
series. As a by-product, the elastic differential cross section
was obtained for the diffraction cone at low and high energies
with a qualitative description of the experimental data. Contrary
to \cite{gay}, in our model we performed a simultaneous fit to the
$\sigma _{tot}$ and $d\sigma /dt$ data as follows.

\begin{figure}[tbp]
\begin{center}
\includegraphics*[scale=0.5]{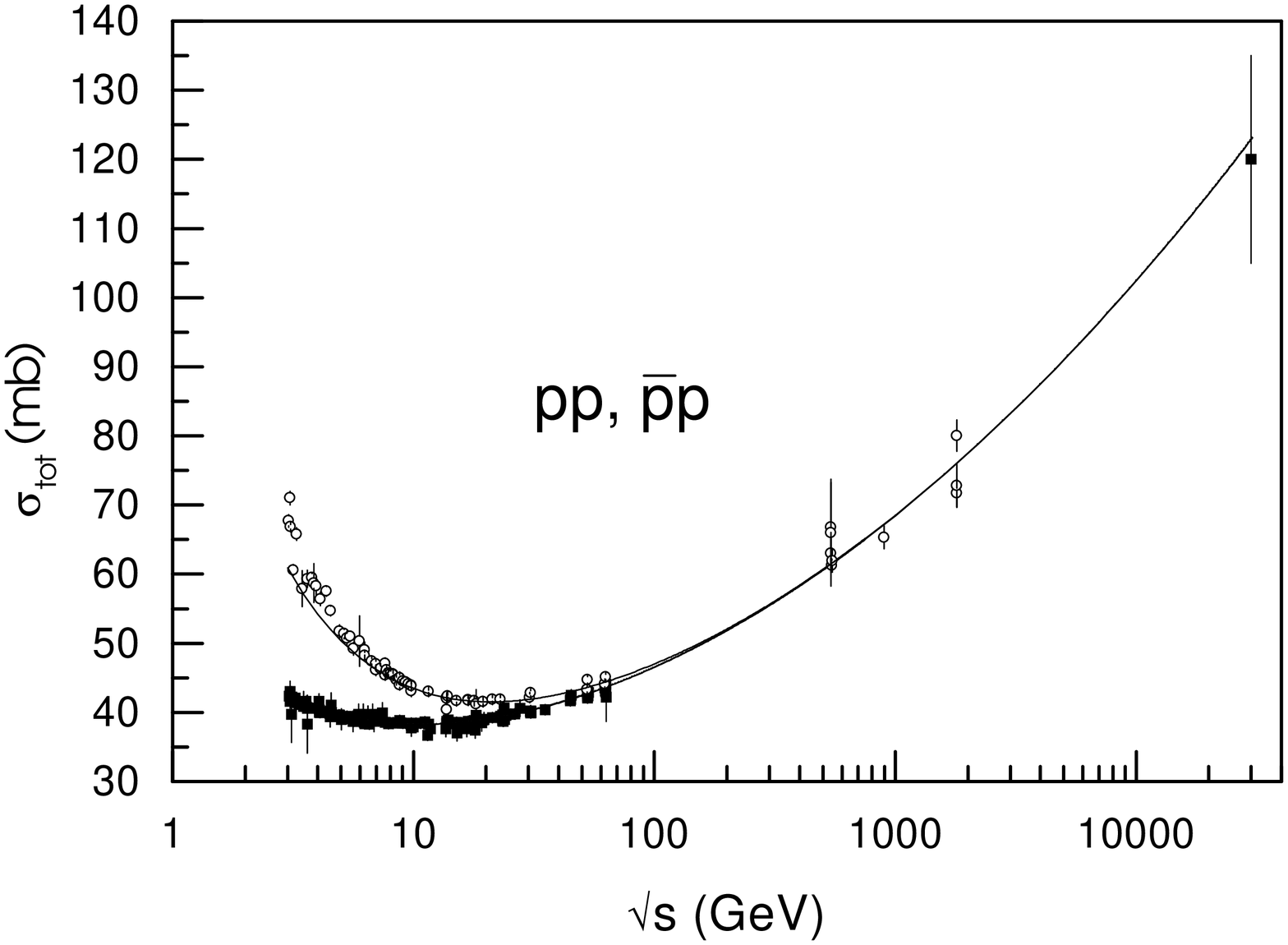}
\caption[]{\small Predictions of the model with the parameters
from Table 1 compared to the experimental data on $\sigma _{tot}$}
\includegraphics*[scale=0.5]{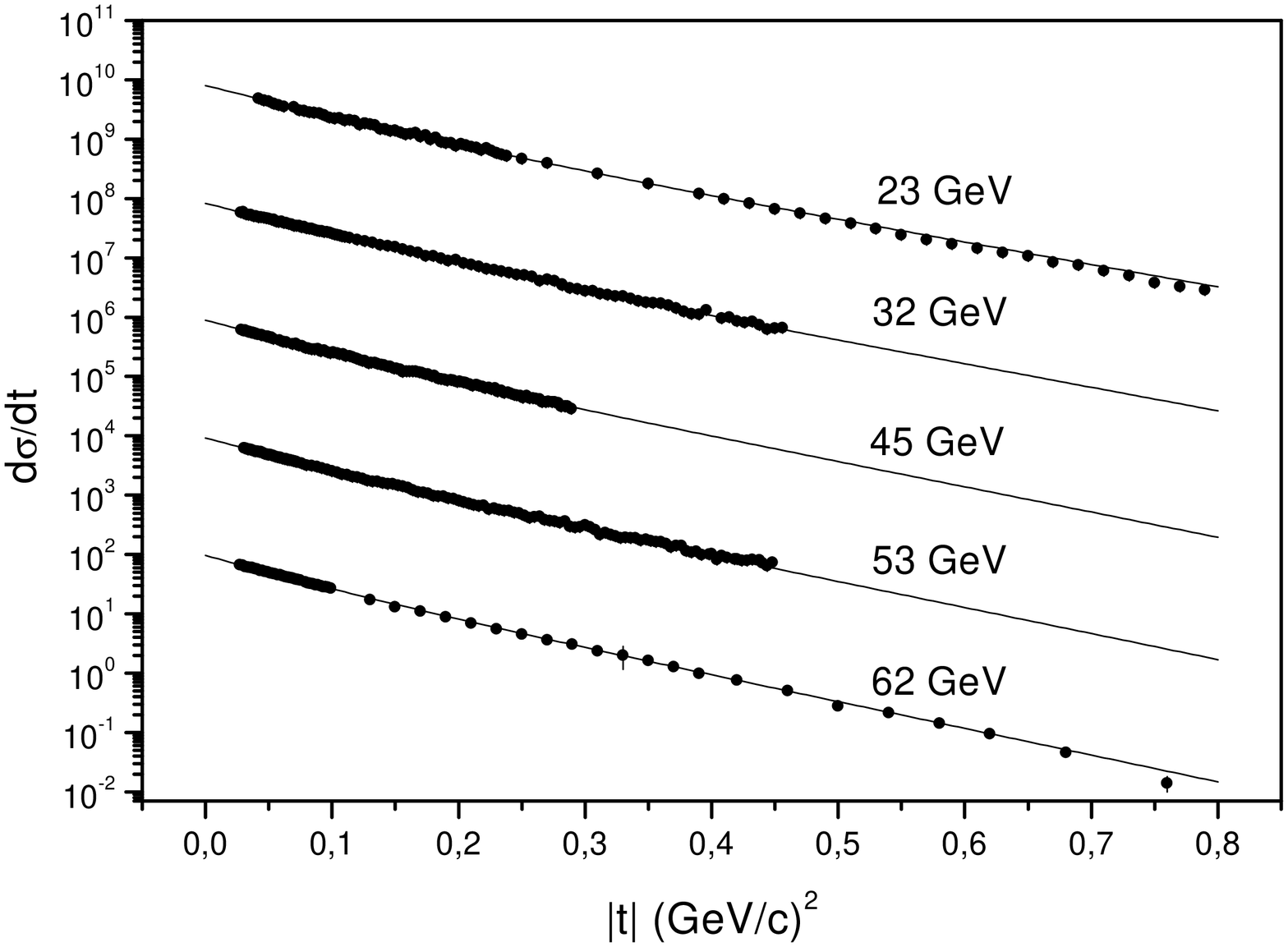}
\caption[]{\small Diffraction cone of $pp-$scattering (a factor
$10^{-2}$ between successive curves is present). The solid curves
are fits of the model.}
\end{center}
\end{figure}

\begin{figure}[tbp]
\begin{center}
\includegraphics*[scale=0.5]{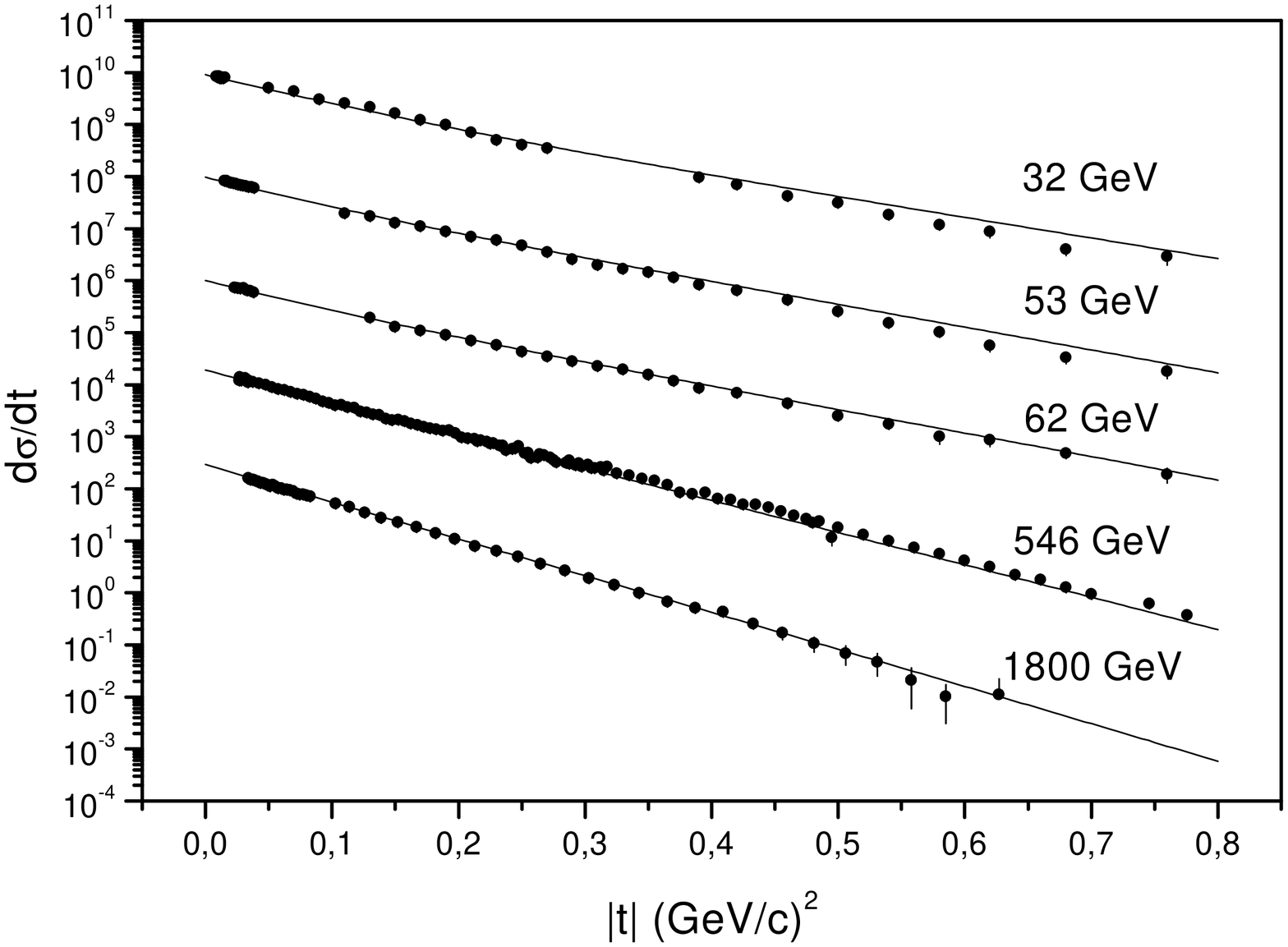}
\caption[]{\small The same as in the Fig. 3 for
$\overline{p}p-$scattering.}
\includegraphics*[scale=0.5]{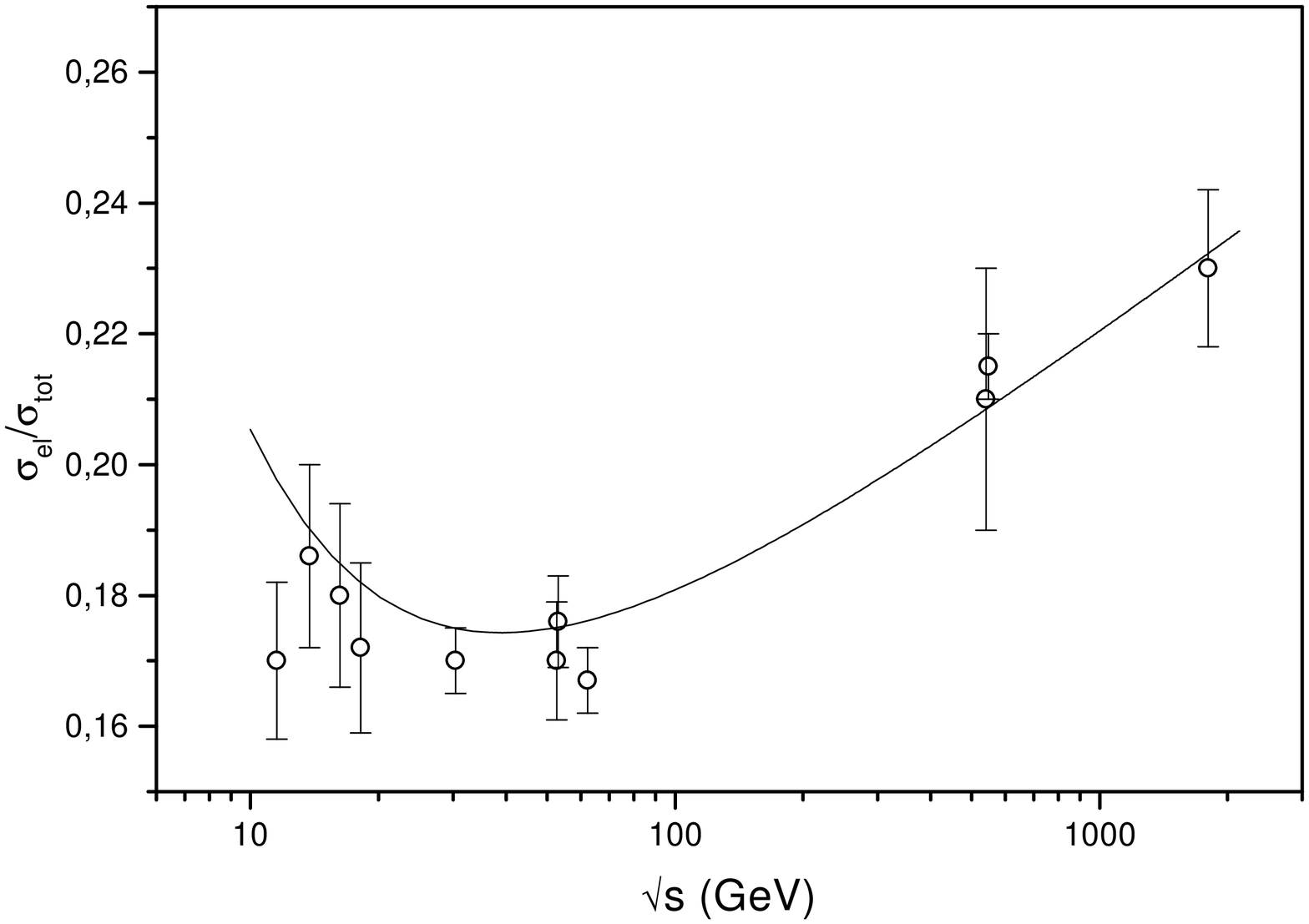}
\caption[]{\small Calculated ratio of $\sigma _{el}/\sigma _{tot}$
for $\overline{p}p-$scattering compared with experiment.}
\end{center}
\end{figure}

\section{Comparison with experiment. Conclusion}

In order to determine the parameters that control the s-dependence
of $A\left(s,0\right)$ in a wide energy range $4GeV\leq
\sqrt{s}\leq 1800GeV$, we used the available data for total cross
sections \cite{SPD}-\cite{AB}. A total of 66 experimental data has
been included for $t=0$. For the differential cross sections we
selected the data at the energies $\sqrt{s}=19;23;31;44;53;62GeV$
(for $pp-$scattering) and $\sqrt{s}=31;53;62;546;1800GeV$ (for
$\overline{p}p-$scattering). The squared 4-momentum has been
limited by $0.05GeV^2<|t|<0.5GeV^2$, because at larger $|t|$ the
influence of the dip region becomes visible (in particular, it can
be seen quite clearly at the Collider energy $\sqrt{s}=546GeV$).
The total of 729 experimental points have been used in the overall
fit.

In the calculations we use the following normalization for the
dimensionless amplitude: $$\sigma _{tot}=\frac{4\pi}sImA\left(
s,t=0\right) ,\frac{d\sigma }{dt}=\frac \pi {s^2}\left| A\left(
s,t\right) \right| ^2.$$

The resulting fits for $\sigma _t$, $\frac{d\sigma }{dt}$ are
shown in Figs. 1-3 with the values of the fitted parameters quoted
in Table 1. From these figures we conclude that the multipole
Pomeron  model corresponding to a sum of gluon ladders up to two
rungs fits the data perfectly well in a wide energy region within
the diffraction cone. As result, the model gives a good behavior
for the ratio $\sigma _{el}/\sigma _{tot}$ for
$\overline{p}p-$scattering (see Fig. 4).

The rapid decrease of the coefficients of the first three terms
(approximately as $1:\frac 1{10}:\frac 1{100}$) in the Pomeron
series in (3) provides the fast convergence of the series and
ensures the applicability of this approximation at still much
higher energies.

\begin{center} {\footnotesize
\begin{tabular}{|ccccccccccc|}
\hline \multicolumn{1}{|c|}{$g_{0}$} & \multicolumn{1}{c|}{$b$} &
\multicolumn{1}{c|}{$c$} & \multicolumn{1}{c|}{$\alpha_{P}'$} &
\multicolumn{1}{c|}{$\gamma_{P}$} &
\multicolumn{1}{c|}{$\alpha_{f}(0)$} &
\multicolumn{1}{c|}{$b_{f}$} & \multicolumn{1}{c|}{$g_{f}$} &
\multicolumn{1}{c|}{$\alpha_{\omega}(0)$} &
\multicolumn{1}{c|}{$b_{\omega}$} & \multicolumn{1}{c|}{$g_{f}$}
\\
\multicolumn{1}{|c|}{} & \multicolumn{1}{|c|}{$(GeV^{-2})$} &
\multicolumn{1}{|c|}{} & \multicolumn{1}{|c|}{$(GeV^{-2})$} &
\multicolumn{1}{|c|}{$(GeV^{-1})$} & \multicolumn{1}{|c|}{} &
\multicolumn{1}{|c|}{$(GeV^{-2})$} & \multicolumn{1}{|c|}{} &
\multicolumn{1}{|c|}{} & \multicolumn{1}{|c|}{$(GeV^{-2})$} &
\multicolumn{1}{|c|}{}
\\ \hline \multicolumn{1}{|c|}{} & \multicolumn{1}{|c|}{} &
\multicolumn{1}{|c|}{} & \multicolumn{1}{|c|}{} &
\multicolumn{1}{|c|}{} & \multicolumn{1}{|c|}{} &
\multicolumn{1}{|c|}{} & \multicolumn{1}{|c|}{} &
\multicolumn{1}{|c|}{} & \multicolumn{1}{|c|}{} &
\multicolumn{1}{|c|}{}
\\
\multicolumn{1}{|c|}{$0.253$} & \multicolumn{1}{c|}{$7.46$} &
\multicolumn{1}{c|}{$0.180$} & \multicolumn{1}{c|}{$0.266$} &
\multicolumn{1}{c|}{$0.137$} & \multicolumn{1}{c|}{$0.777$} &
\multicolumn{1}{|c|}{$3.04$} & \multicolumn{1}{c|}{$13.3$} &
\multicolumn{1}{c|}{$0.524$} & \multicolumn{1}{c|}{$6.86$} &
\multicolumn{1}{c|}{$7.97$}
\\
\multicolumn{1}{|c|}{} &
\multicolumn{1}{|c|}{} & \multicolumn{1}{|c|}{} &
\multicolumn{1}{|c|}{} & \multicolumn{1}{|c|}{} &
\multicolumn{1}{|c|}{} & \multicolumn{1}{|c|}{} &
\multicolumn{1}{|c|}{} & \multicolumn{1}{|c|}{} &
\multicolumn{1}{|c|}{} & \multicolumn{1}{|c|}{}
\\
\hline
\end{tabular}
}
\end{center}

\begin{center}
{\small Table 1: Values of the fitted parameters in the model.}
\end{center}

In this paper we have explored only the simplest extension of the
dipole Pomeron to a tripole. In fact, the scattering amplitude is
much more complicated than just a simple power series in $\ln s$.

Earlier a comprehensive analysis of the $pp-$ and $\overline{p}p-$
diffraction cone scattering using the dipole and supercritical
Pomeron models was done \cite{DLM}.

On the one hand, though we used just a simplified t-dependence in
the model reasonably good results were obtained. Because the
slopes of secondary Reggeons do not influence the fit
sufficiently, we have fixed them at $\alpha_{f}'=0.84$
$(GeV^{-2})$ and $\alpha_{\omega}'=0.93$ $(GeV^{-2})$, which
correspond to the values of Chew-Frautschi plot. On the other
hand, we included the curvature of the Pomeron trajectory that
cannot be negligible. It should be also taken into account that
the calculated slope of the Pomeron value is close to the
generally accepted value $\alpha_{P}'=0.25$ $(GeV^{-2})$.

The quality of our fits $\chi ^2/dof=2.25$ has not reached the
best fit $\chi ^2/dof=1.3$ obtained in \cite{DLM}. Nevertheless,
we believe this difference can be significantly reduced after
adding the data of Coulomb interference region $|t|<0.05GeV^2$ to
the global fit.

\bigskip

{\large \bf Acknowledgments} We thank L. Csernai, L. Jenkovszky
and F. Paccanoni for discussions and BCPL, Bergen University,
where this paper was completed, for hospitality and financial
support. The work was supported also by the Ukrainian-Hungarian
grant 2M/200-2000.

\end{document}